\documentclass{ws-ijmpe}

\begin{document}

\markboth{P.C. Srivastava and I. Mehrotra}{Nuclear structure study around $Z=28$}

\catchline{}{}{}{}{}
%%%%%%%%%%%%%%%%%%%%%%%%%%%%%%%%%%%%%%%%%%%%%%%%%%%%%%%%%%%%%%%%%%%%

\title{Nuclear structure study around $Z=28$}

\author{P.C. SRIVASTAVA\footnote {praveen.srivastava@nucleares.unam.mx}}

\address{Department of Physics, University of Allahabad, Allahabad-211002, India,}
\address{Grand Acc\'el\'erateur National d'Ions Lourds (GANIL), CEA/DSM--CNRS/IN2P3, BP~55027, F-14076 Caen Cedex 5, France, and}
\address{Instituto de Ciencias Nucleares, Universidad Nacional Aut\'onoma de M\'exico, 04510 M\'exico, D.F., M\'exico } 
  
\author{I. MEHROTRA}

\address{Department of Physics, University of Allahabad, Allahabad-211002, India}

\maketitle

%\pub{Received (Day Month Year)}{Revised (Day Month Year)}

\begin{abstract}

Yrast levels of Ni, Cu and Zn isotopes for $40 \leq N
\leq50$ have been described by state-of-the-art shell model calculations with 
three recently available
interactions using $^{56}$Ni as a
core in the $f_{5/2}pg_{9/2}$ model space. The results
are unsatisfactory
viz. large $E(2^+)$ for very neutron rich nuclei,
small $B(E2)$ values in comparison to experimental values.
 These results indicate an importance of inclusion
of $\pi f_{7/2}$ and $\nu d_{5/2}$ orbitals in the model space 
to reproduce collectivity in this region. 

\end{abstract}

\ccode{PACS Nos.: 21.60.Cs, 27.50.+e.}

%=================================================================

\section {Introduction}
 \label{s_intro} 

 The neutron-rich nuclei near nickel region is one of the fascinating subject because many intensive experimental investigations have been done in the last few years. \cite{Sor02,Fra98,Fra01,Walle07,Stone08,Kenn02,Grz05,Chiara11} 
The reason for this interest is due to the evolution of the single-particle structure towards $^{78}$Ni which is a testing ground for the nuclear shell model and importance of astrophysical $\it{r}$ process, which is the mechanism of rapid neutron capture by seed nuclei in explosive stellar environments. \cite{Kra93} A hitherto question in this region related to rapid reduction in the energy of 5/2$^-$ state
as the filling of neutrons started in the $\nu g_{9/2}$ orbital in the Cu isotopes. \cite{Ste08,Ste09,Ily09,Dau10,Fla09,Fla10} The importance of monopole term from the tensor force is pointed out by Otsuka {\it et al.}, \cite{Ots05,Ots10} to understand the evolution of nuclear structure in this region. 
Collectivity, and $B(E2)$ enhancement around $N=40$ for Cr, Fe and Ni isotopes recently studied by including $\pi f_{7/2}$ and $\nu d_{5/2}$ orbitals in the model space. \cite{Cau02,Lenzi10,Lju10,Roth11} It is found that for neutron rich $fp$ shell nuclei the deformation is appears in these nuclei due to coupling of unfilled $f_{7/2}$ proton shell to neutron in $sdg$ shell.

\begin{figure}
\begin{center}
\resizebox{90mm}{45mm}{\includegraphics{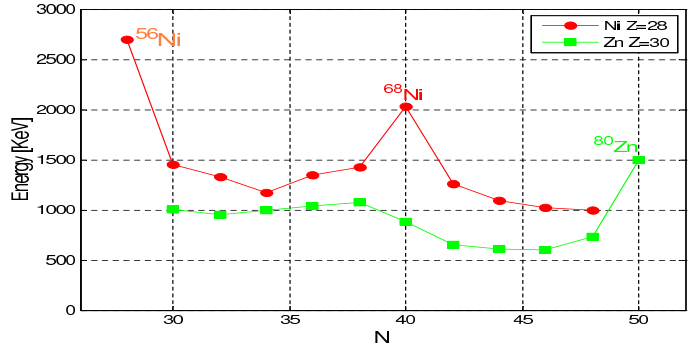}}
\end{center}
\caption{Systematic of the experimentally observed $E(2_1^+)$ in the stable and neutron-rich Ni and Zn isotopes near the $N~=~40$ and 50 shell closure.}
\end{figure}

The experimental $E(2_1^+)$ for Ni and Zn isotopes are shown in Fig. 1. In this figure $E(2_1^+)$ states in Zn are overall lower compared to Ni and an additional decrease of the $E(2_1^+)$ energy in Zn isotopes is obtained between $N=40-50$ compared to $N=28-40$. This is probably due to increased $\pi$-$\nu$ interaction between the two protons outside the $Z=28$ shell and the $N=40-50$ neutron shell, which bring in an amount of collectivity.

 Following our recent shell-model (SM) studies for
neutron-rich F isotopes \cite{Sri11}, odd and even isotopes of Fe
\cite{Sri09,Sria}, odd-odd Mn isotopes \cite{Srib}, 
odd-mass $^{61,63,65}$Co isotopes \cite{Sric}, and odd-even Ga isotopes \cite{Srid}, in the present work, large scale shell model calculations have been performed for neutron rich Ni, Cu and Zn isotopes for $40 \leq N \leq50$ in $f_{5/2}pg_{9/2}$ model space. The low-lying energy levels and $B(E2)$ values have been calculated and compared with the recent experimental data.

The paper is organized as follows: In Section 2 model space and effective interaction are described.
In Section 3 results and discussion are presented. Finally in Section 4 we give conclusions.

%=================================================================     
\section{Model space and effective interaction}

 The calculations have been performed using $p_{3/2}$, $f_{5/2}$, $p_{1/2}$ and $g_{9/2}$ valence space taking $^{56}$Ni as a core. In the present calculation we use three different sets of interactions. The first set of large scale shell model calculations (labeled LSSMI) utilizes the realistic effective nucleon-nucleon interaction based on G-matrix theory by Hjorth-Jensen \cite{Jen95} with the monopole modification by Nowacki  \cite{Now96,Smi04}. The second set of calculations (labeled LSSMII) are obtained with the JJ44B effective interaction \cite{Lis04} which is an extension of the renormalized G-matrix interaction based on the Bonn-C {\it NN} potential (JJ4APN) constructed to reproduce the experimental data for exotic Ni, Cu, Zn, Ge and {\it N=50} intones in the vicinity of $^{78}$Ni. The third set of calculations (labeled LSSMIII) is performed with JUN45 interaction due to Honma {\it et al.} \cite{Honma09}\\ 
The single particle energies for the first set of calculation (LSSMI) for orbital $p_{3/2}$, $f_{5/2}$, $p_{1/2}$ and $g_{9/2}$ are 0.000, 0.770, 1.113 and 3.000 MeV respectively. These single particle energies are taken from experimental data of $^{57}$Ni.  The single-particle energies for the second set of calculations (LSSMII) for the orbital $p_{3/2}$, $f_{5/2}$, $p_{1/2}$,and $g_{9/2}$ are -9.65660, -9.28590, -8.26950 and -5.89440 MeV respectively. 
The single-particle energies for JUN45 interaction (labeled LSSMIII)
are taken to be -9.8280, -8.7087, -7.8388, and -6.2617 MeV
for the $p_{3/2}$, $f_{5/2}$, $p_{1/2}$ and $g_{9/2}$
orbital, respectively.
All the SM calculations have been carried out using the
code {\tt ANTOINE }~\cite{Now96,Caurier05,Caurier99} at SGI Cluster computer at \textsc{ganil} and KanBalam facility of \textsc{dgctic-unam}.

\begin{figure}
\begin{center}
\resizebox{100mm}{45mm}{\includegraphics{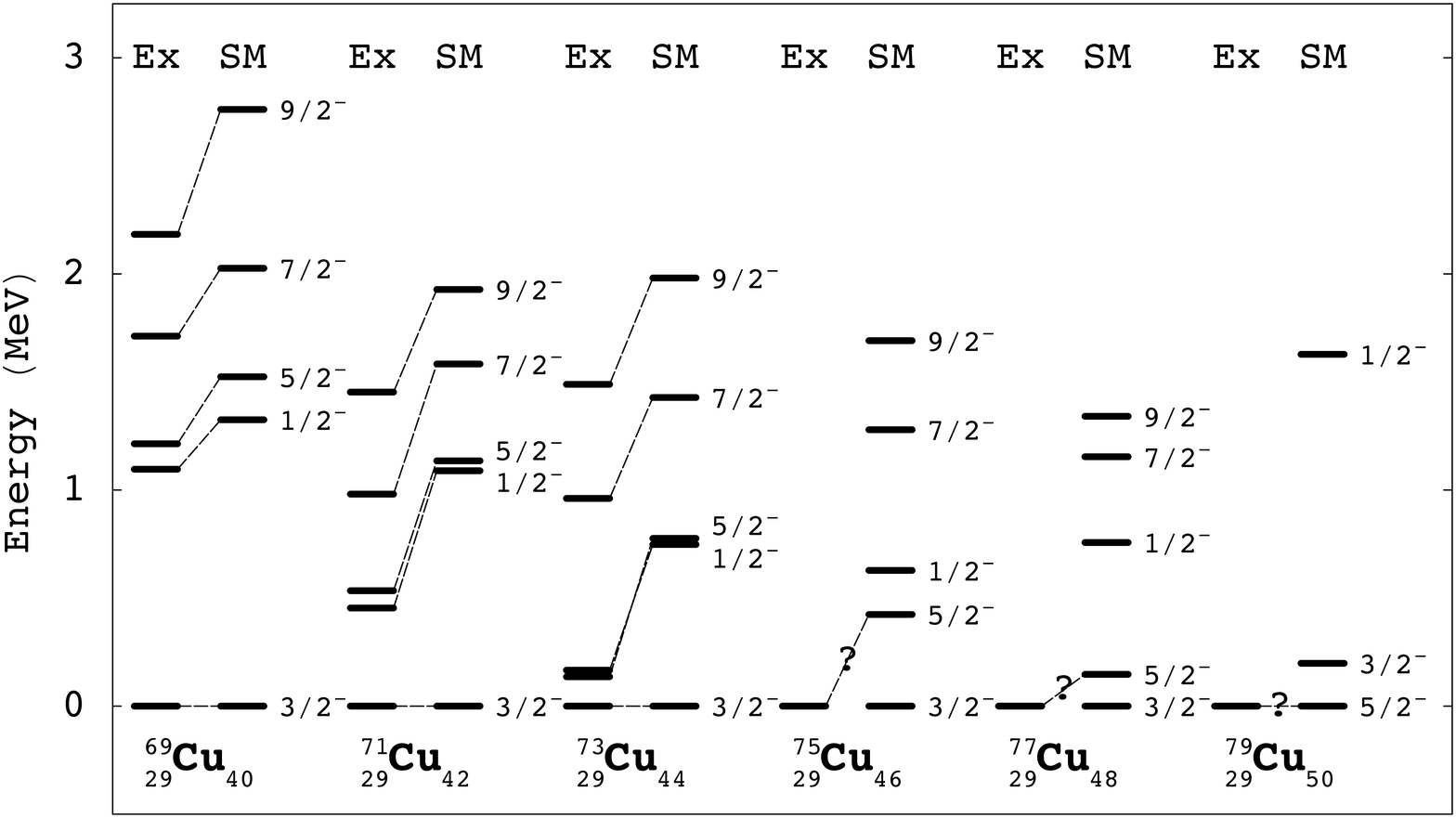}}
\end{center}
\caption{Yrast levels of $^{69-79}$Cu isotopes using LSSMI interaction.}
\end{figure} 

%=================================================================
\section{ Results and Discussion}
    
\subsection{Low lying energy levels of $^{69-79}$Cu}
The yrast levels of $^{69-79}$Cu isotopes for $^{56}$Ni core using these three interactions are shown in Figs. 2, 3 and 4 by indicating Nowacki interaction as (LSSMI), Lisetskiy interaction as (LSSMII) and Honma interaction as (LSSMIII). For $^{69-73}$Cu with  LSSMI ground state spin is correctly predicted but other yrast levels are slightly higher in energy in comparison to the experimental value. For $^{75-77}$Cu, predicted ground state spin is 3/2$^-$ whereas the experimental value is 5/2$^-$. For $^{79}$Cu, predicted spin of 5/2$^-$ agrees with the experimental value.
\begin{figure}[h]
\begin{center}
\resizebox{100mm}{45mm}{\includegraphics{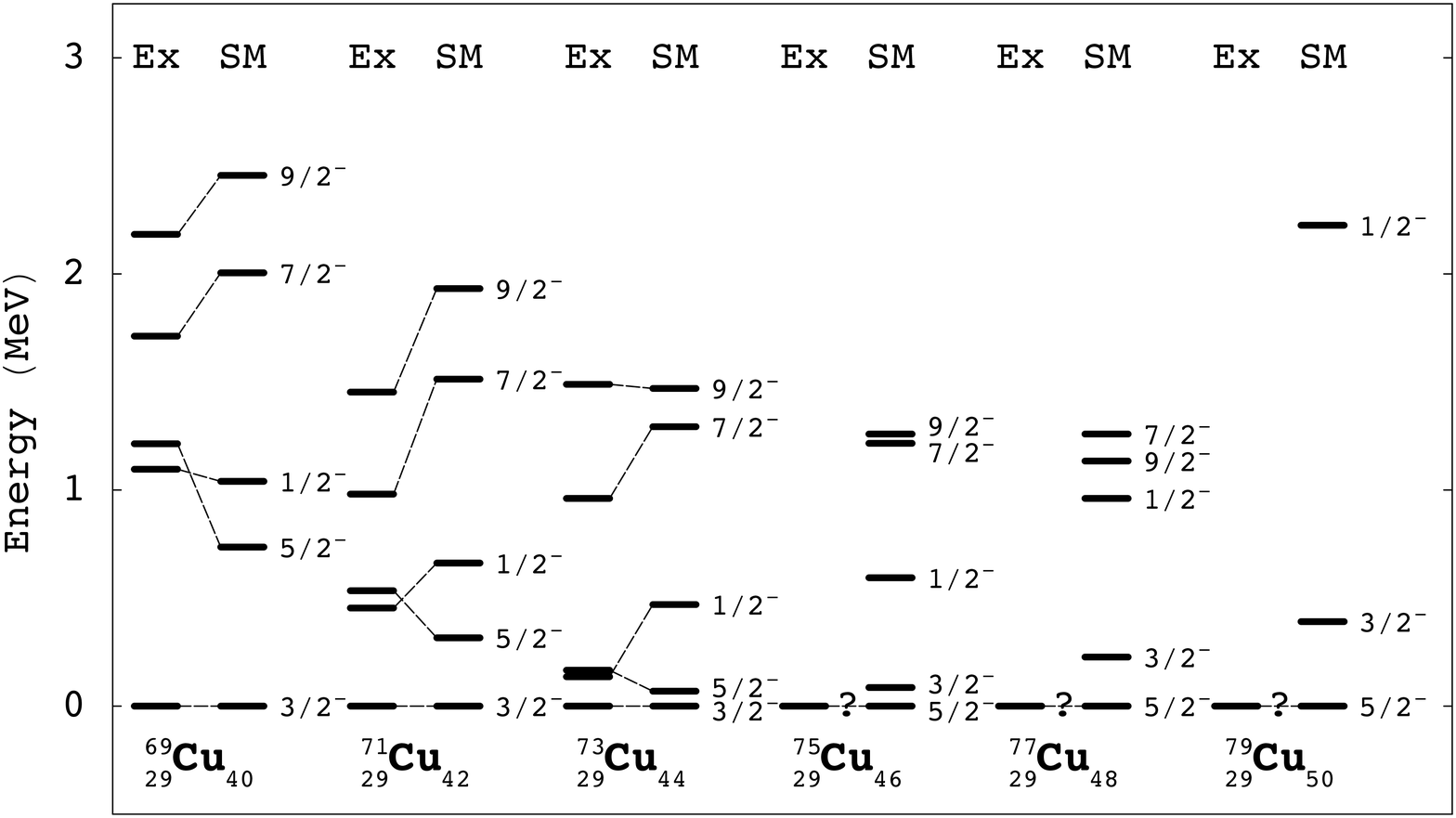}}
\end{center}
\caption{Yrast levels of $^{69-79}$Cu isotopes using LSSMII interaction.}
\end{figure}

\begin{figure}
\begin{center}
\resizebox{125mm}{50mm}{\includegraphics{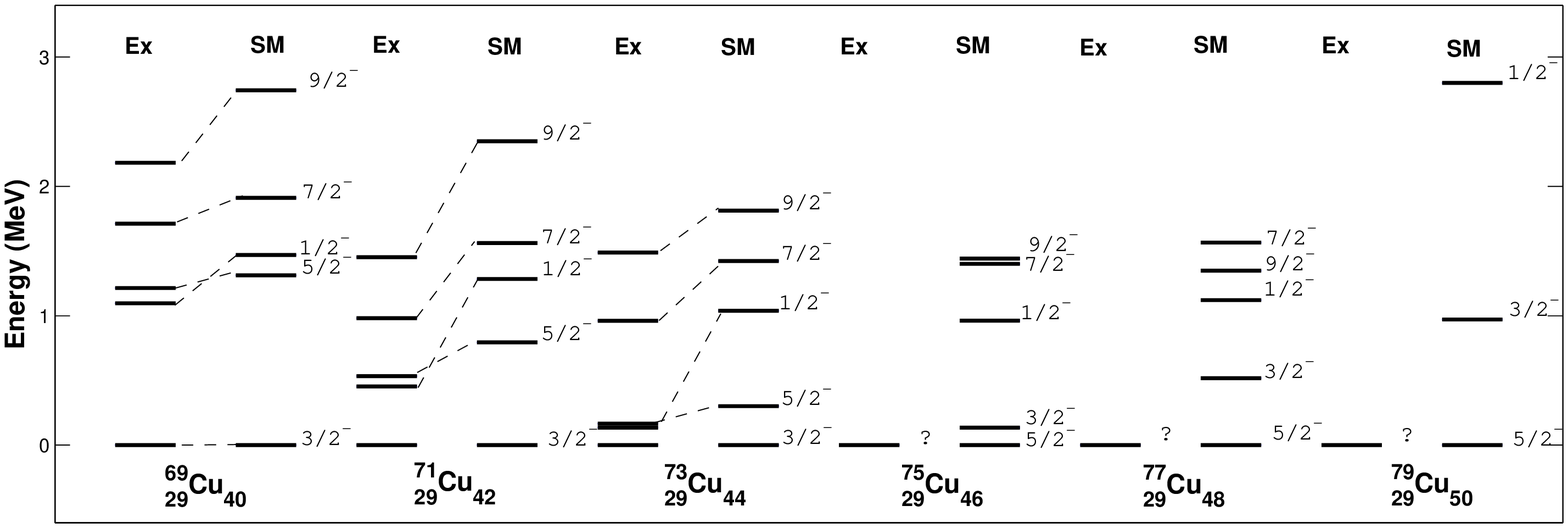}}
\end{center}
\caption{Yrast levels of $^{69-79}$Cu isotopes using LSSMIII interaction.}
\end{figure}

 The LSSMII calculations predict correct ground state for all the $^{69-79}$Cu isotopes. For $^{69}$Cu and $^{71}$Cu isotopes yrast levels are compressed  
compared to the experimental value. For  $^{71}$Cu only first 5/2$^-$ is lower in energy while the other levels are higher in energy in comparison to the corresponding experimental values. It is observed that in going from $^{69}$Cu to $^{79}$Cu, as the neutron number increases, the first 5/2$^-$ gets lower  and lower in energy and becomes the ground state for $^{75}$Cu onwards. It is observed that in going from $^{69}$Cu to $^{73}$Cu energy gap between first excited 5/2$^-$ state and 3/2$^-$ ground state decreases in energy. For $^{75}$Cu the two levels cross each other and 5/2$^-$ state becomes the  ground state. The gap again rises for $^{79}$Cu. The 1/2$^-$ predicted by this interaction is high in $^{73,75}$Cu.

The LSSMIII calculations predict correct ground state for all the $^{69-79}$Cu isotopes. The position of the first 5/2$^-$ level is resonable in comparision to LSSMI and LSSMII. With this interaction first 1/2$^-$ is too high ( $\sim$ 800 keV).
This is probably due to missing $\pi f_{7/2}$ orbital in the model space. The spacing between the 5/2$^-$ and the 3/2$^-$ start increasing beyond $^{75}$Cu.
It is important to investigate onset of collectivity beyond $N=40$ by including  $\pi f_{7/2}$ and $\nu d_{5/2}$ orbitals in the model space.

\subsection{Low lying energy levels of $^{68-76}$Ni}
The low-lying energy levels of $^{68-76}$Ni isotopes are shown in Figs. 5, 6 and 7 for three interactions.  The order of energy levels are well reproduced for both LSSMI, LSSMII and LSSMIII interactions. For $^{68}$Ni, the first 4$^+$ state is predicted slightly lower in energy by LSSMI. The LSSMII predicts 2$^+$ and 4$^+$ states about at 319 keV and at 277 keV above the experimental value.  For $^{70}$Ni the LSSMI and LSSMII interactions predict slightly higher energy of the first 2$^+$ state. The LSSMI and LSSMII predict 2$^+$  state above the experimental value by 406 keV and 172 keV. As the number of neutron increases from $^{72}$Ni onwards both the interactions predict higher levels nearly 500 keV higher than the experimental values. Thus both these interaction fail to reproduce the experimental data of more neutron rich Ni isotopes.

\begin{figure}
\begin{center}
\resizebox{125mm}{45mm}{\includegraphics{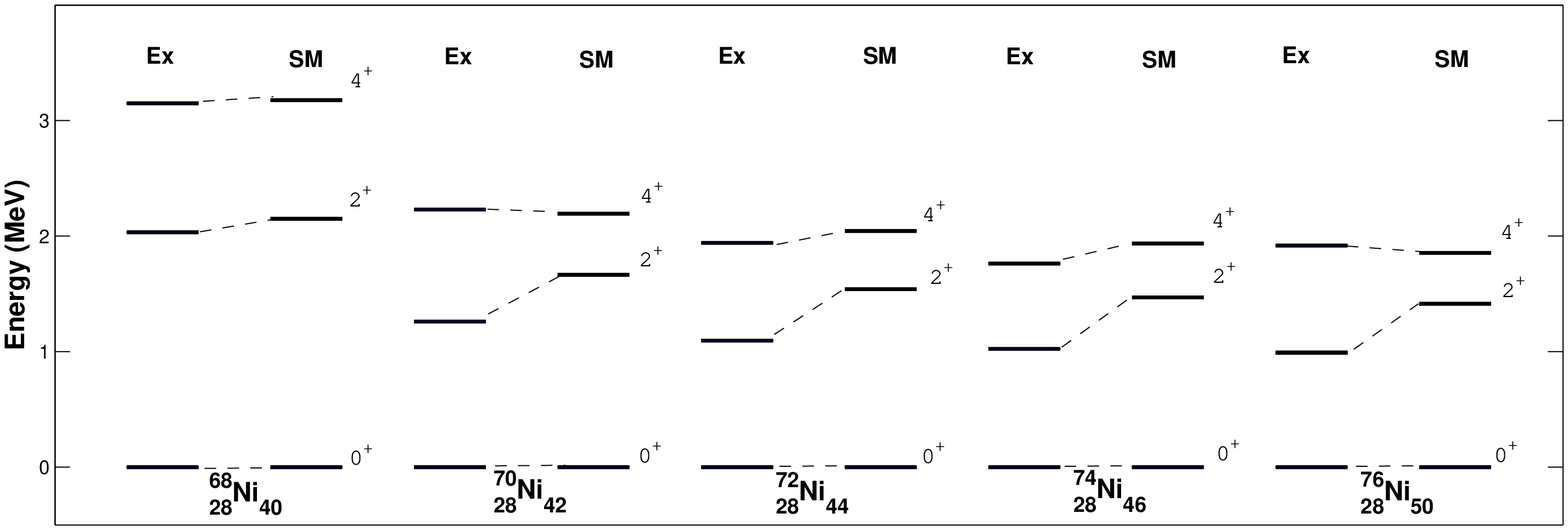}}
\caption{Yrast levels of $^{68-76}$Ni isotopes using LSSMI interaction.}
\end{center}
\end{figure}
\begin{figure}
\begin{center}
\resizebox{125mm}{45mm}{\includegraphics{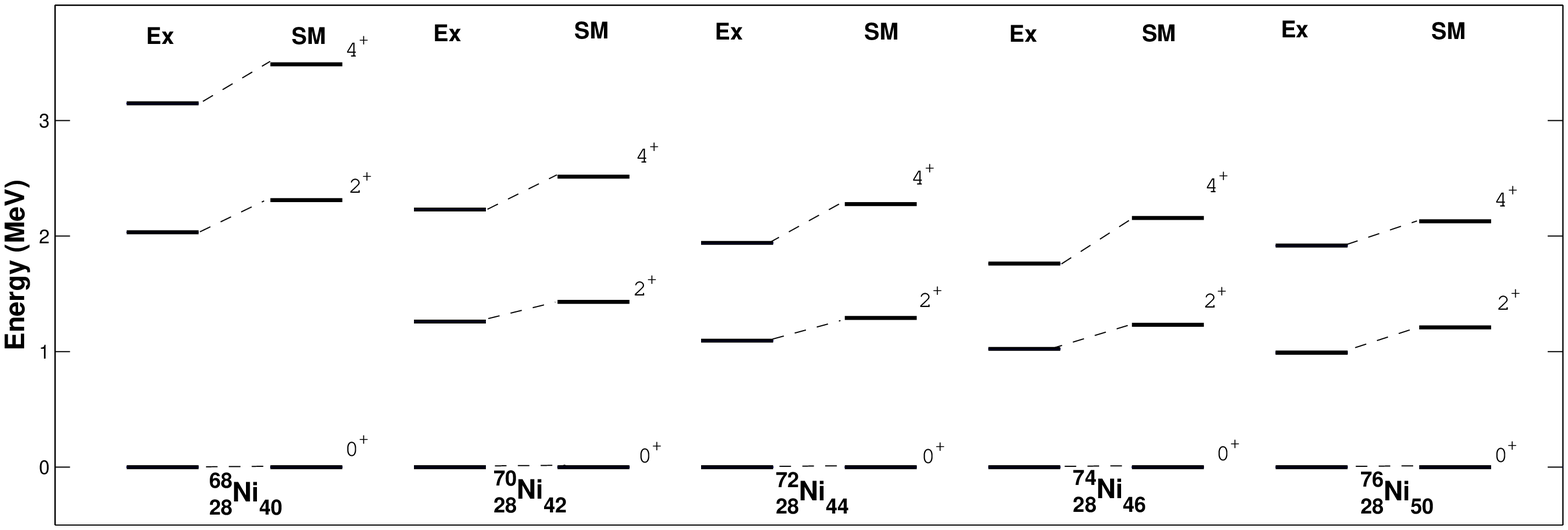}}
\caption{Yrast levels of $^{68-76}$Ni isotopes using LSSMII interaction.}
\end{center}
\end{figure}
\begin{figure}
\begin{center}
\resizebox{125mm}{45mm}{\includegraphics{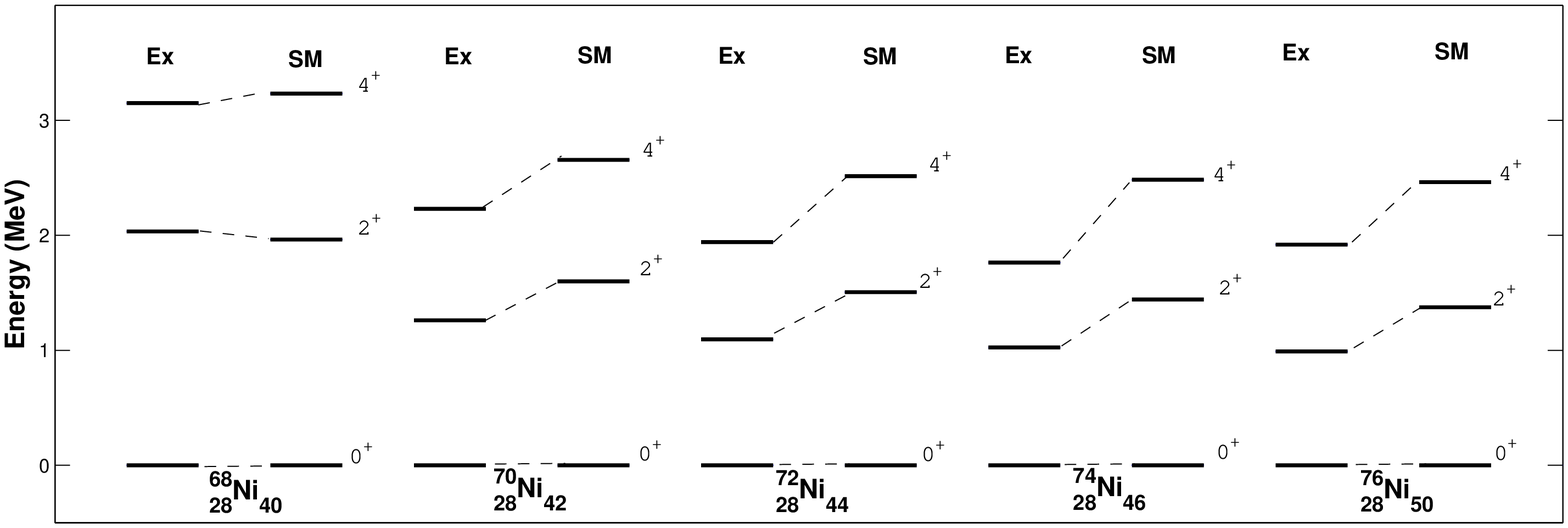}}
\caption{Yrast levels of $^{68-76}$Ni isotopes using LSSMIII interaction.}
\end{center}
\end{figure}

\begin{figure}
\begin{center}
\resizebox{120mm}{!}{\includegraphics{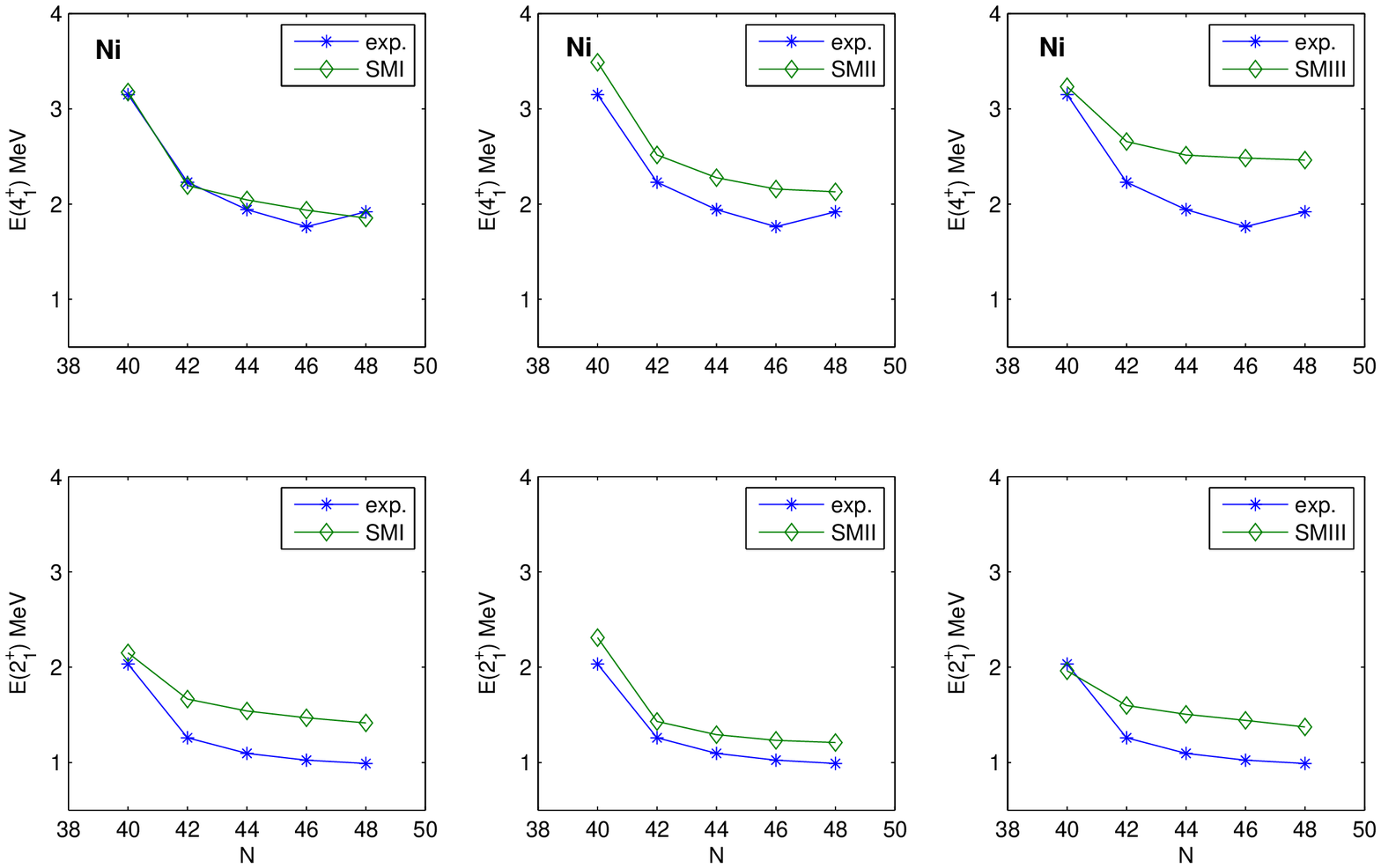}}
\end{center}
\caption{ The calculated and experimental $E(2_1^+)$ and $E(4_1^+)$ for Ni  isotopes as a function of neutron number for LSSMI, LSSMII and LSSMIII interactions.}
\end{figure}

\begin{figure}
\begin{center}
\resizebox{125mm}{45mm}{\includegraphics{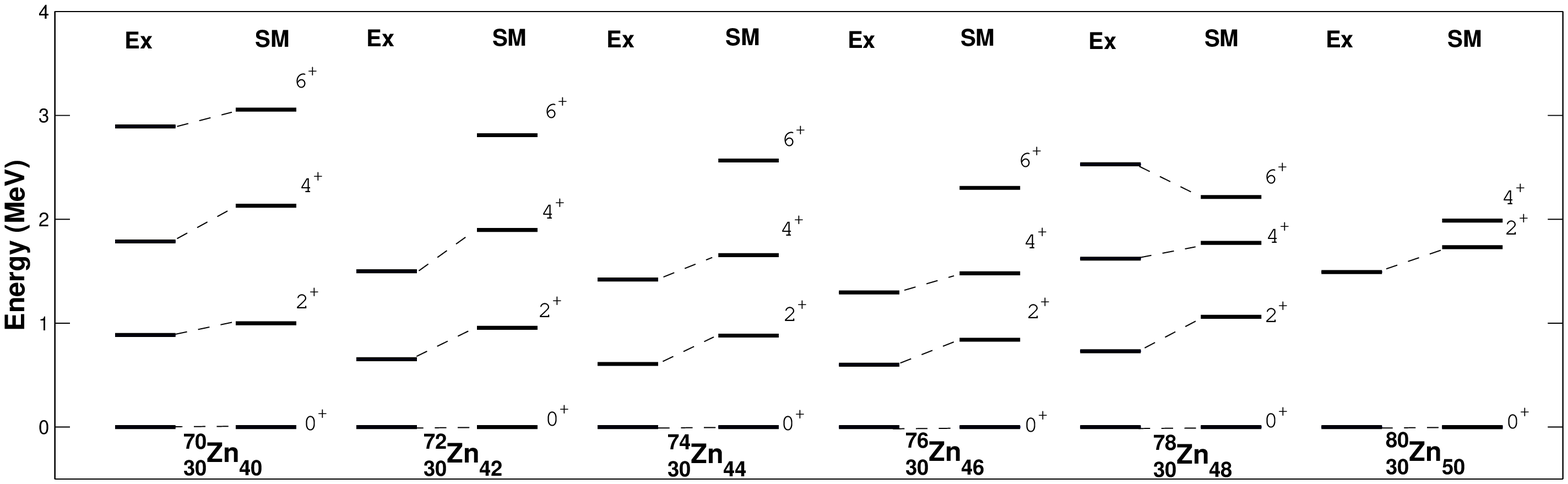}}
\caption{Yrast levels of $^{70-80}$Zn isotopes using LSSMI interaction.}
\end{center}
\end{figure}

\begin{figure}
\begin{center}
\resizebox{125mm}{45mm}{\includegraphics{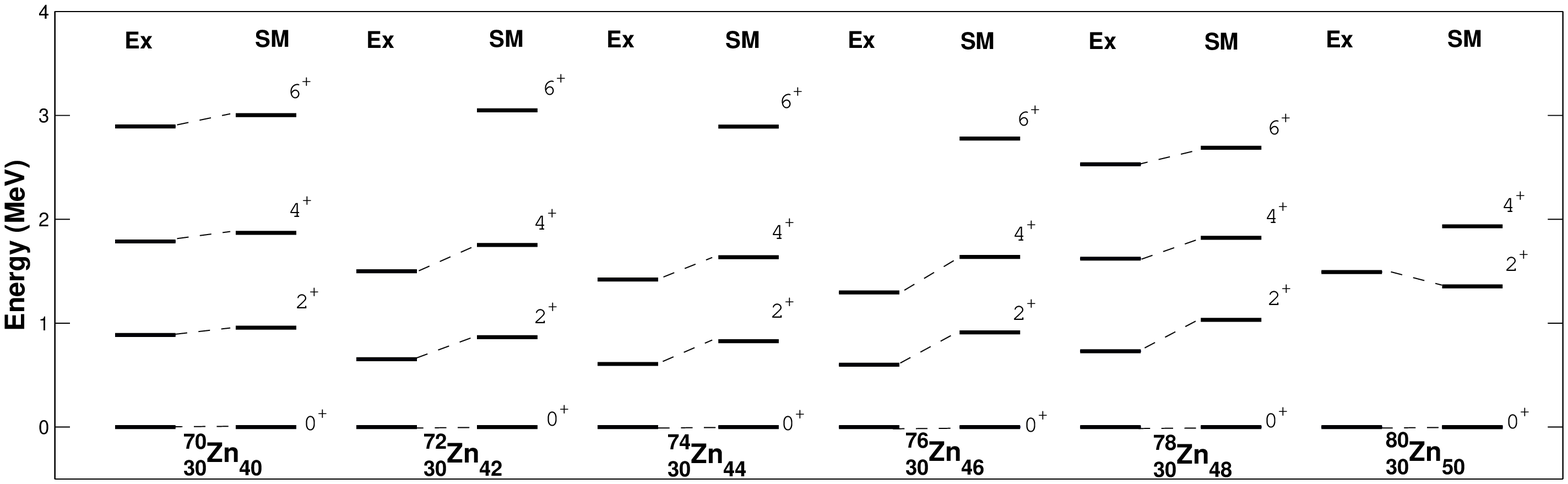}}
\caption{Yrast levels of $^{70-80}$Zn isotopes using LSSMII interaction.}
\end{center}
\end{figure}

\begin{figure}
\begin{center}
\resizebox{125mm}{45mm}{\includegraphics{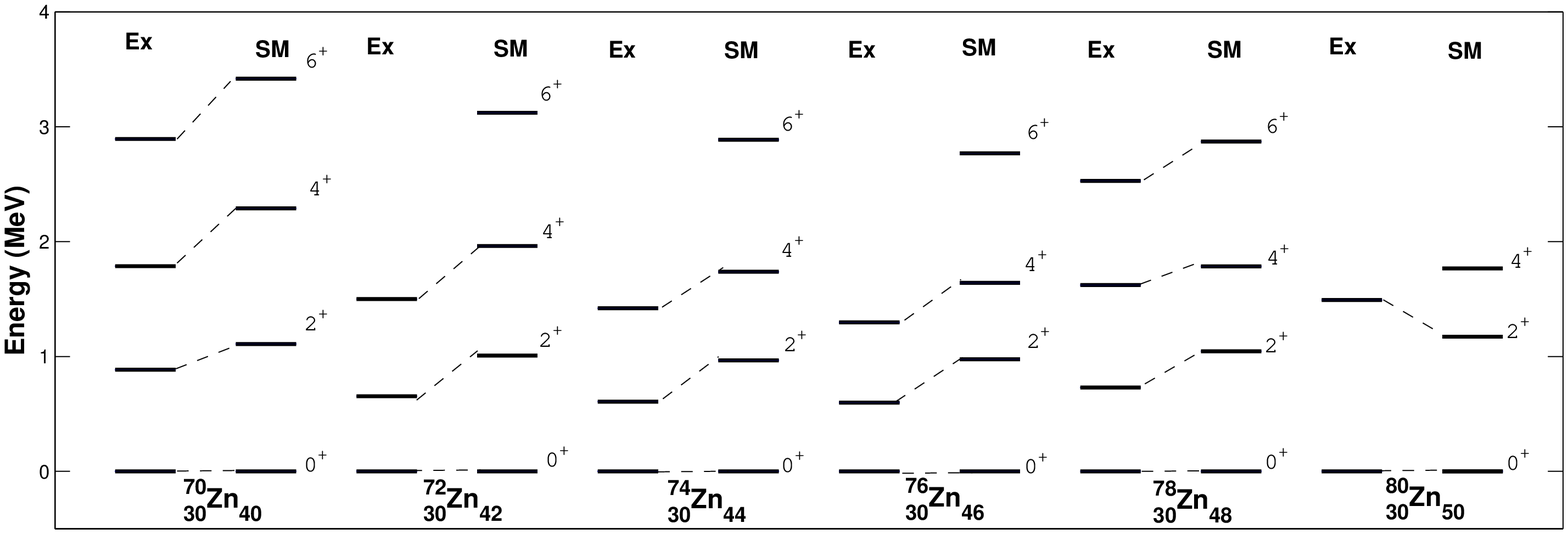}}
\caption{Yrast levels of $^{70-80}$Zn isotopes using LSSMIII interaction.}
\end{center}
\end{figure}

The LSSMIII, for $^{68}$Ni predict  2$_1^+$ lower in comparison to experimental data, while calculated 4$_1^+$ is closed to experimental data. This interaction predict higher values of excitation energies for 2$_1^+$ and 4$_1^+$ beyond $N=40$.

 The calculated and experimental values of $E(2_1^+)$ and $E(4_1^+)$ for Ni  isotopes are shown in Fig. 8. The $E(4_1^+)$ energy is well predicted by LSSMI and $E(2_1^+)$ by LSSMII. The high value of $E(2_1^+)$ at $N~=~40$ is a clear indication of shell closure. 
The discrepancy between calculated and experimental value is due to insufficient quadrupole collectivty. Indeed, to reproduce collectivity for Ni isotopes it is necessory to include $f_{7/2}$ and $d_{5/2}$ single-particle orbits in the model space. Recently collectivity for Ni isotopes using this space is reported by 
modern effective interaction by Strasbourg group.\cite{Cau02,Lenzi10}

%\newpage
\begin{figure}
\begin{center}
\resizebox{120mm}{!}{\includegraphics{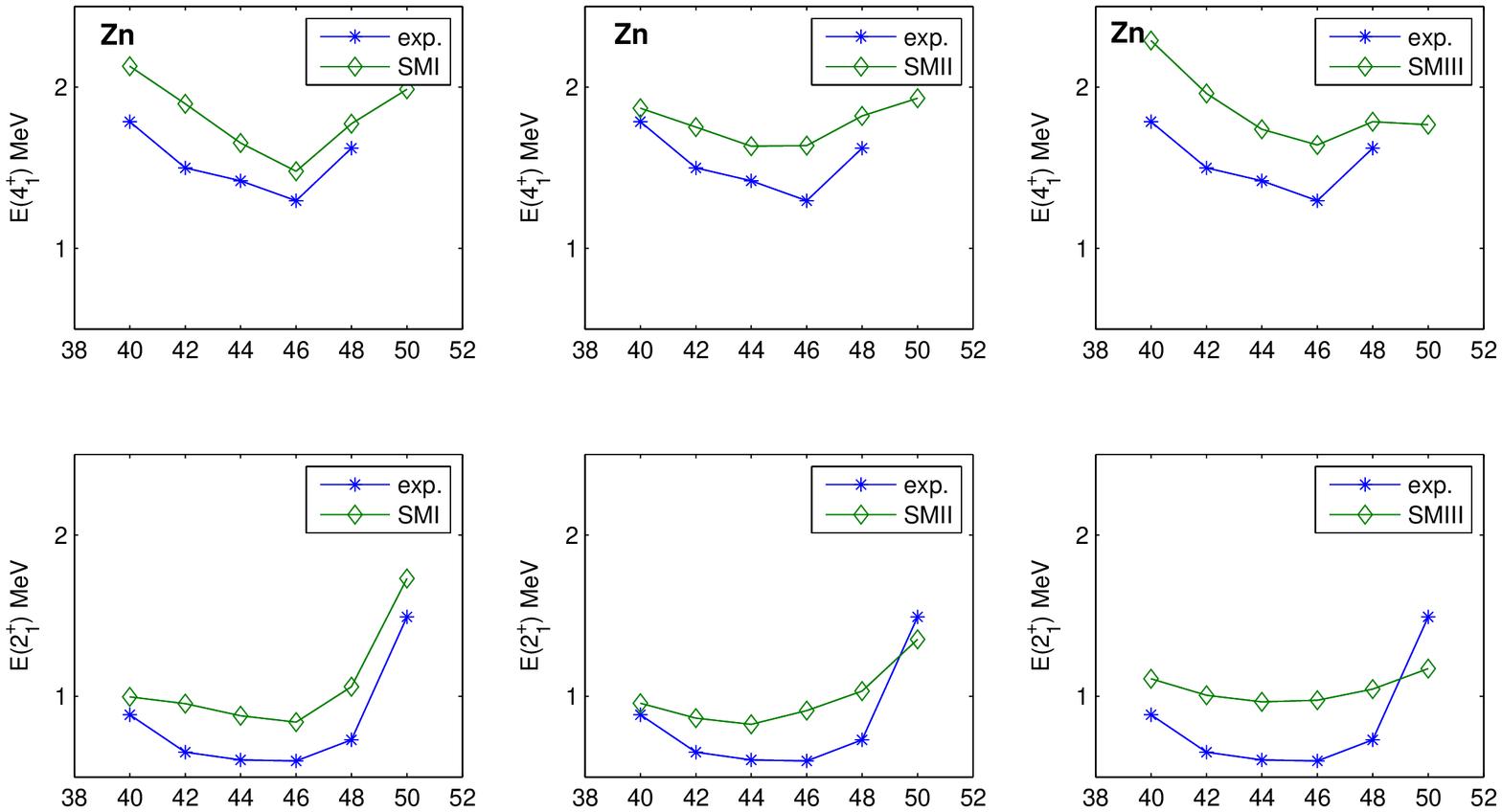}}
\end{center}
\caption{ The calculated and experimental $E(2_1^+)$ and $E(4_1^+)$for Zn  isotopes as a function of neutron number for LSSMI, LSSMII and LSSMIII interactions.}
\end{figure}

\begin{figure}
\begin{center}
\resizebox{110mm}{90mm}{\includegraphics{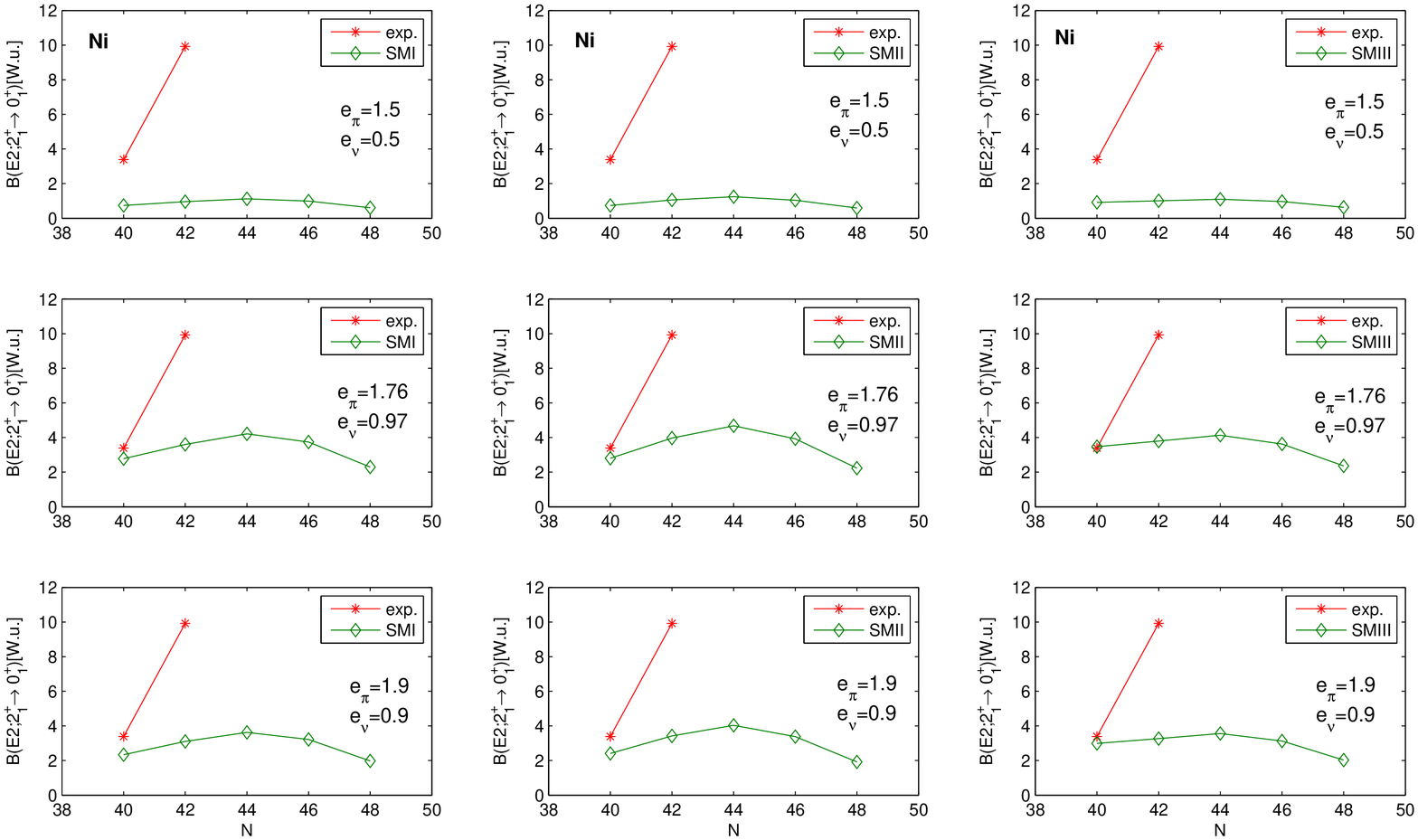}}
\end{center}
\caption{ The calculated and experimental $B(E2;2_1^+\rightarrow 0_{gs}^+)$ values for Ni  isotopes as a function of neutron number for LSSMI, LSSMII and LSSMIII interactions.}
\end{figure}

\begin{figure}
\begin{center}
\resizebox{110mm}{90mm}{\includegraphics{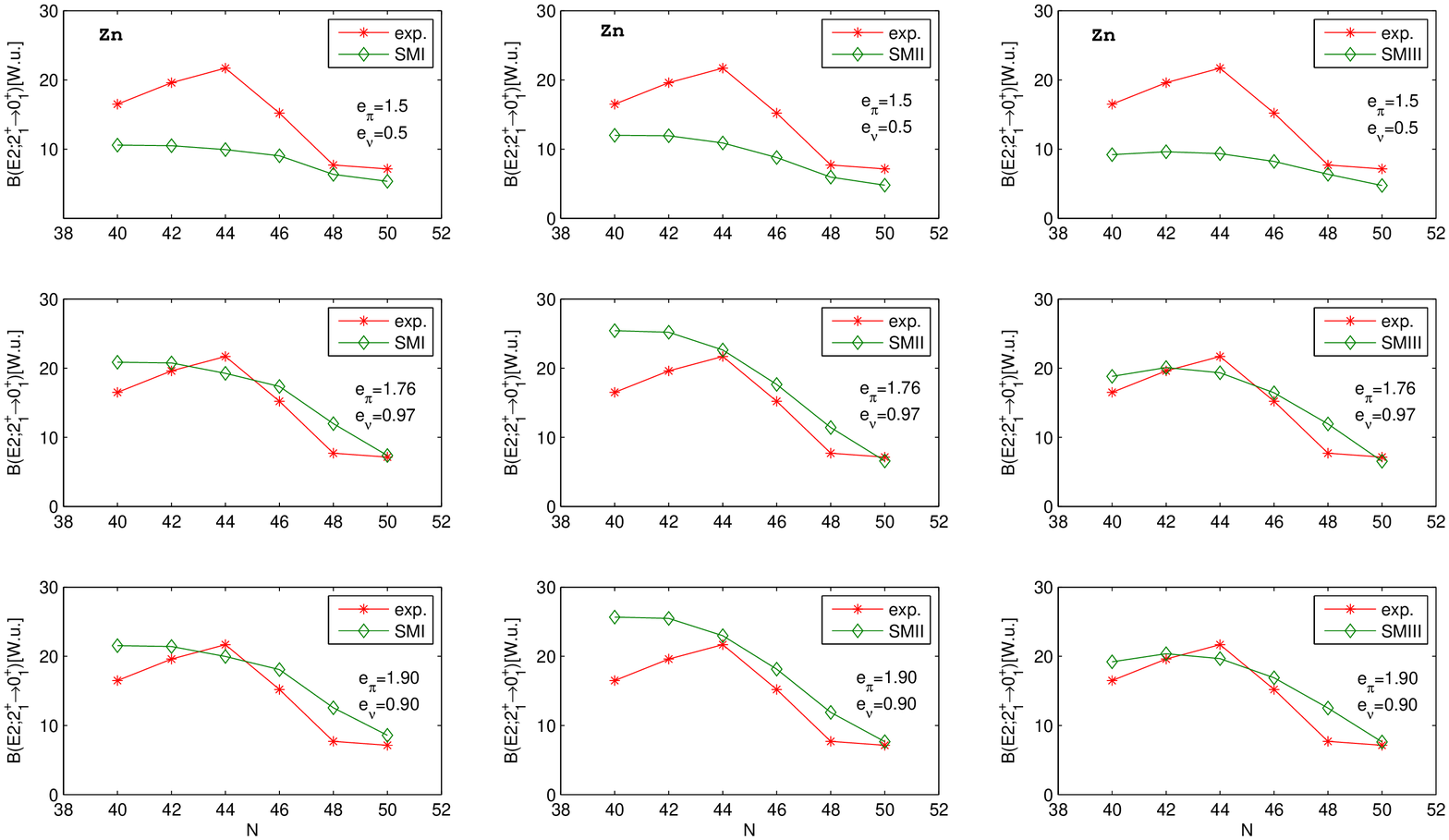}}
\end{center}
\caption{ The calculated and experimental $B(E2;2_1^+\rightarrow 0_{gs}^+)$  values for Zn isotopes as a function of neutron number for LSSMI, LSSMII
and LSSMIII interactions.}
\end{figure}

\subsection{Low lying energy levels of $^{70-80}$Zn}
The low-lying energy levels of $^{70-80}$Zn isotopes are shown in Figs. 9, 10 and 11. For $^{70}$Zn the three interactions predict correct order of levels. The first 2$^+$ state at 885 keV are predicted at 997 keV and at 957 keV by LSSMI and LSSMII interactions respectively. The LSSMII predicts better results for the first  2$^+$ state. The second  0$^+$ state is predicted 500 keV higher than the experimental value. For $^{72}$Zn, the order of levels is correctly reproduced by LSSMII. The predicted 2$^+$ state at 302 keV and at 211 keV above the experimental value for LSSMI and LSSMII interactions respectively. For $^{74}$Zn, 2$^+$ is predicted at 879 keV and at 826 keV by LSSMI and LSSMII whereas the experimental at 606 keV. The results of LSSMII are better than LSSMI. For $^{76}$Zn, the 2$^+$ at 599 keV is predicted at 840 keV and at 911 keV by LSSMI and LSSMII respectively. The 2$^+$  in $^{78}$Zn at 730 keV is predicted at 1060 and 1032 keV and in $^{80}$Zn at 1492 keV is predicted at 1731 keV and 1353 keV for LSSMI and LSSMII interactions. Thus as number of neutron increases both these interactions give higher values of 2$^+$ state in comparison to the corresponding experimental values. The LSSMIII  calculations predict large discrepancies with experimental data.

For Zn isotopes, the calculated and experimental values of $E(2_1^+)$ and $E(4_1^+)$  for both the interactions are shown in Fig. 12. The $E(2_1^+)$ is predicted correctly by LSSMII interaction. The large value of $E(2_1^+)$ at $N~=~50$ is a clear indication of shell closure at $N~=~50$. 
For Zn isotopes it is important to investigate onset of collectivity beyond $N=40$ by including  $\pi f_{7/2}$ and $\nu d_{5/2}$ orbitals in the model space.

\subsection{The $B(E2)$ systematics in the Ni isotopes}
 $B(E2; 2_1^+\rightarrow$0$_{gs}^+)$ values for Ni isotopes for different set of effective charges for LSSMI, LSSMII and LSSMIII are shown in Fig. 13. For $^{68}$Ni, the $B(E2)$ for $e_\pi$=1.76 and $e_\nu$=0.97 is more closer to experimental values in comparison to the other set of values. The high values of $e_\pi$  are required to reproduce experimental results indicating a strong $Z=28$ core polarization. The experimental $B(E2; 2_1^+\rightarrow$0$_{gs}^+)$ in $^{70}$Ni is not reproduced within $f_{5/2}p_{9/2}$ space. 
This has been interpreted in Ref. \cite{Per06} as a rapid polarization of the proton core induced by the filling of the neutron $g_{9/2}$ orbit. This reflects a strong monopole interaction between $\pi1f_{7/2}$-$\nu1g_{9/2}$. These results show that the dominance of $g_{9/2}$ orbit in $B(E2)$ calculation is important above $N~=~40$. Recently Lenzi {\it et al.},\cite{Lenzi10} obtained good agreement for $B(E2; 2_1^+\rightarrow$0$_{gs}^+)$ with experimental data in Ni isotopes by including $\pi f_{7/2}$ and $\nu d_{5/2}$ orbitals in the model space.

\subsection{The $B(E2)$ systematics in the Zn isotopes}
Recently Coulomb excitation experiment have been performed at \textsc{rex-isolde}, to measure $ B(E2; 2_1^+\rightarrow$0$_{gs}^+)$  values in $^{74-80}$Zn, $ B(E4; 4_1^+\rightarrow$2$_{1}^+)$  values in $^{74,76}$Zn and to determine first excited $2_1^+$ in $^{78,80}$Zn.\cite{Walle09} 
The $ B(E2; 2_1^+\rightarrow$0$_{gs}^+)$ values for Zn isotopes for different set of effective charges for LSSMI, LSSMII and LSSMIII are shown in Fig. 14. This figure show that higher value of $e_\pi$ is required to reproduce $B(E2)$ correctly. 
In Zn isotopes above A=68, the contribution from $\nu 1g_{9/2}^2$ configuration in the wave function of the low lying excited $2_1^+$ and $4_1^+$ states is important. The $B(E2; 2_1^+\rightarrow$0$_{gs}^+)$ strength is thus dominated by the specific $E2$ strength between ($\nu 1g_{9/2}$)$_{J=0}$ and ($\nu 1g_{9/2}$)$_{J=2}$  configurations. The increase of the $B(E2)$ values beyond $N~=~40$ suggests an increase of the collectivity induced by the interaction of protons in the {\it pf} shell and the neutrons in $sdg$ orbitals.

%=================================================================
%\newpage
\section{Conclusions}
The large $E(2_1^+)$ values for Ni and Zn isotopes and corresponding low $B(E2)$ values indicate that the inclusion of $\pi f_{7/2}$ and $\nu d_{5/2}$ orbitals in the model space is important. The increase of the $B(E2)$ values beyond $N~=~40$ suggests an increase of the collectivity induced by the interaction of protons in the $pf$ shell and neutron in the $sdg$ shell. Recently \textsc{lnps} interaction for $fpgd$ space due to Lenzi {\it et al.},\cite{Lenzi10}  have been reported in the literature. This interaction account collectivity toward $N~=~40$ for Cr, Fe and Ni isotopes. 

%-------------------------------------------------------------------------
\section*{Acknowledgments}

Thanks are due to Prof. P.~Van Isacker for his suggestions during the work.  All the calculations in the present paper are carried out using the \textsc{sgi} cluster resource at \textsc{ganil} and \textsc{dgctic-unam} computational facility KanBalam. This work was financially supported by the Sandwich PhD programme of the Embassy of France in India. We acknowledge fruitfull discussions on collectivity with M. J. Ermamatov.

%-------------------------------------------------------------------------

\end{document}